\documentclass[twocolumn,twoside]{article}
\usepackage{graphics}
\usepackage{amsmath,amssymb}

\newcommand{\zav}[1]{\left(#1\right)}
\newcommand{\hzav}[1]{\left[#1\right]}

\newcommand{\hb}{\\ \hspace*{2ex}}

\begin{document}
\title{The benefits of the orthogonal LSM models}
\author{Z.\,Mikul\'a\v{s}ek$^{1,2}$\\[2mm] 
\begin{tabular}{l}
 $^1$ Institute for Theoretical Physics and Astrophysics, Masaryk
 University\hb  Kotl\'a\v{r}sk\'a 2, CZ-611 37 Brno, Czech Republic,
 {\em mikulas@physics.muni.cz}\\
 $^2$ Observatory and Planetarium of J. Palisa, V\v{S}B-Technical
 University\hb
 Ostrava, Czech Republic\\[2mm]
\end{tabular}
}
\date{}
\maketitle

ABSTRACT. In the last few decades both the volume of high-quality
observing data on variable stars and common access to them have
boomed; however the standard used methods of data processing and
interpretation have lagged behind this progress. The most popular
method of data treatment remains for many decades Linear Regression
(LR) based on the principles of Least Squares Method (LSM) or
linearized LSM. Unfortunately, we have to state that the method of
linear regression is not as a rule used accordingly namely in the
evaluation of uncertainties of the LR parameters and estimates of
the uncertainty of the LR predictions.

We present the matrix version of basic relations of LR and the true
estimate of the uncertainty of the LR predictions. We define
properties of the orthogonal LR models and show how to transform
general LR models into orthogonal ones. We give relations for
orthogonal models for common polynomial series.\\[1mm]


{\bf 1. Introduction}\\[1mm]

The development in the field of variable stars research from
Tsessevich's times is enormous. The number of known variable stars
has arisen by at least two orders, as well as the number of their
observers and interpreters. It has arisen both the volume and common
access to high-quality variable stars observing data and
computational techniques. The number of new efficient statistical
techniques and methods that are available for everybody thanks to
wide spread personal computers have been developed and published.
Nevertheless, the methods used for processing of variable stars data
mostly have remained the same as those used in Vladimir
Platonovich's era.

Every astrophysicist likes large quantities and better quality of
modern observational data, new methods of processing are not so
popular. Majority of them needs a good knowledge of matrix calculus,
what is in discordance with a frequent syndrome of variable stars
observers, which could be named \emph{Matrixphobia}. Very rarely we
are encountering with the opposite syndrome of \emph{Matrixphilia}
which invades mathematically erudite theoreticians loving new
methods and matrices so much that they do not use them for real
observational data. Both extremes in the data processing are bad and
we should find our golden mean.

The contemporary statistics shares inexhaustible quantity of
methods. It is necessary to select several of the most versatile and
diverse methods, master them and to learn to combine them. The
method of processing must not be unique, but always must be
made-to-measure of the set problem.

The majority of variable stars data processing tasks are solved
using least square method, strictly speaking linear regression,
where as models serve the most frequently common polynomials or
sine/cosine series. It should be noted that there exist several
other methods which are able to give the same or better results. One
of them is for example the Advanced Principal Component Analysis,
which is the combination of LSM and standard Principal Component
Analysis (see Mikul\'a\v{s}ek, 2007). The method is optimal for
solving of a lot astrophysics problems as a realistic fitting of
multicolour light curves, the determination of the moments of
extrema of multicolour light curves, modeling of light multicolour
curves which is necessary for the process of improvement of
ephemerides, diagnostics of light curve (LC) secular changes, and
the classification of LCs. Other methods of modern data treatment
are also mentioned in Andronov, I., these Proceedings.

In the following section we will pay attention to some details of
linear regression procedure which is very likely the most frequently
used tool of variable stars data processing.
\\[2mm]

{\bf 2. The Least Squares Method}\\[1mm]

The very frequent astrophysical task is to fit a curve through a
series of $N$ observed points described by a triad
$\{x_i,y_i,w_i\}$, where $x_i$ is an independent (well measured)
quantity like time or a phase, related to the $i$-th measurement
$y_i$ is the dependent, measured quantity like magnitude, O$-$C, and
$w_i$ is the weight of the measurement, as a rule inversely
proportional to the square of the expected uncertainty of the value
$y_i$. Hereafter we will use normalized weights $w_i$ the mean value
$\bar{w}$ of which is equal to 1.

$F(x,\vec{\beta})$ is so called \emph{model function} of $x$
described by the $k$ free parameters
$\beta_1,\beta_2,\ldots,\beta_k$ arranged into the vector
$\vec{\beta}$. We define a function of this vector $S(\vec{\beta})$:
\begin{equation}
S(\vec{\beta})=\sum_{i=1}^N
\hzav{\,y_i-F(x_i,\vec{\beta})}^2\,w_i.\label{S1}
\end{equation}
The solution of the LSM minimalization procedure, is finding of the
vector of parameters $\vec{\beta}=\mathbf{b}$, for which is the
quantity $S(\vec{\beta})$ minimal. The success of the method in the
given situation depends above all on our skill in the creating of
the mathematical model expressed by the function $F(x,\vec{\beta})$.
Then the finding of the best fit in the range of functions
admissible by the pre-selected model is relatively simple and
straightforward. In principle it is solution of $k$ equations of $k$
unknown parameters arranged in the vector $\mathbf b$:
\begin{equation}
\left. \frac{\partial S}{\partial
\vec{\beta}}\right|_{\vec{\beta}=\mathbf{b}}=
\mathbf{grad}\!\hzav{S(\vec{\beta}=\mathbf{b})}=\vec{0}.\
\Rightarrow
\end{equation}
\begin{equation}
\sum_{i=1}^N y_i\,\frac{\partial F(x_i,\mathbf{b})}{\partial
\beta_j}\,w_i=\sum_{i=1}^N F(x_i,\mathbf{b})\,\frac{\partial
F(x_i,\mathbf{b})}{\partial \beta_j}\,w_i,\label{S3}
\end{equation}
for $j=1,2,\ldots,k$.\\[2mm]

{\it 2.1. Linear regression}\\[1mm]

The LSM procedure of the determination of the solution will be
considerably simplified if we use the linear model of the found
function $F(x,\vec{\beta})$, assuming:
\begin{equation}
F(x,\vec{\beta})=\sum_{j=1}^k \beta_j\,f_j(x),\\
\end{equation}
where $f_j(x)$ are arbitrary functions of $x$. Eq.\,\ref{S1} then
can be rewritten in the form:
\begin{equation}
S(\vec{\beta})=\sum_{i=1}^N \hzav{\,y_i-\sum_{j=1}^k
\beta_j\,f_j(x_i) }^2\,w_i. \label{LF}
\end{equation}
 Eq.\,\ref{S3} then switches to:
\begin{equation}
\sum_{i=1}^N y_i\,f_j(x_i)\,w_i=\sum_{i=1}^N \hzav{\,\sum_{p=1}^k
b_p\,f_p(x_i)}f_j(x_i)\,w_i,\label{LR1}
\end{equation}
It is advantageous to express all operations in matrix form. Then
\begin{equation}
\mathbf{X}=\left(
    \begin{array}{cccc}
      f_1(x_1) & f_2(x_1) & \cdots & f_k(x_1) \\
      f_1(x_2) & f_2(x_2) & \cdots  & f_k(x_2) \\
      \vdots & \vdots & \ddots & \vdots \\
      f_1(x_N) & f_2(x_N) & \cdots  & f_k(x_N) \\
    \end{array}
  \right),
\end{equation}
\begin{equation}
\mathbf{Y}=\zav{y_1\,y_2\,\cdots\,y_N}^{\mathrm{T}};\,
\mathbf{W}=\mathbf{diag}\zav{w_1\,w_2\,\cdots\,w_N},
\end{equation}
\begin{gather}
\mathbf{H={\zav{X^T W\,X}}^{-1}},\ \mathbf{b=H\,X^T W\,Y},\
\mathbf{Y_p=X\,b},\label{def1}\\
R=\mathbf{Y^T W\,Y - b^T X^T W\,Y},\ s=\sqrt{\frac{R}{(N-k)}},
\end{gather}
where $\mathbf{Y}_\mathrm{p}$ is the vector of the predictions,
$\overline{w}$ is the mean value of weights $w_i$, $R$ is the
weighted sum of square deflections, $s$ is the weighted standard
deviation of the fit.

The procedure of linear regression with the explicit linear model is
quick and its solution is unique. In the general case we may find
several solutions although some of them could be physically unreal.
The most common method of finding of local minima on the $S(\beta)$
plane is an iterative gradient method, where we use the above
mentioned apparatus of linear regression applied on the linearized
model function.\\[2mm]

{\it 2.2. Linearized regression}\\[1mm]

The linearization of the general model function $F(x,\vec{\beta})$
consists in substitution of it by its Taylor expansion in respect of
$\vec{\beta}$. We need to know as good as possible estimate
$\mathbf{b_e}$ of the solution of LSM equations $\mathbf{b}$,
$\mathbf{b_e} \rightarrow\mathbf{b}$. Then we can write:
\begin{equation}
 F(x_i,\vec{\beta})\cong F(x_i,\mathbf{b_e})+\sum_{j=1}^k
\frac{\partial F(x_i,\mathbf{b_e})}{\partial \beta_j}
(\beta_j-b_{\mathrm e j}).
\end{equation}
\begin{equation}
S(\vec{\beta})=\sum_{i=1}^N \hzav{\,\Delta y_i-\sum_{j=1}^k
f_j(x_i)\,\Delta \beta_j}^2\,w_i,\label{S2}
\end{equation}
where
\begin{gather}
\Delta y_i=y_i-F(x_i,\mathbf{b_e}),\quad f_j(x)=\frac{\partial
F(x,\mathbf{b_e})}{\partial \beta_j},\nonumber\\
\Delta{\vec{\beta}}=\vec{\beta}-\mathbf{b_e}.
\end{gather}
The equations Eq.\,\ref{LF} and Eq.\,\ref{S2} are formally
identical, despite the meanings of particular terms in them are
different. We define column vector $\Delta\!\mathbf{ Y}=[\Delta
y_1\,\Delta y_2\,\cdots\,y_N]$, and the column vector of the
correction of the solution estimate $\mathbf{b_e}$,
$\Delta{\mathbf{b}}$.
\begin{eqnarray}
\mathbf{H={\zav{X^\mathrm{T} W\,X}}}^{-1},\
\Delta\mathbf{b=H\,X^\mathrm{T} W} \Delta\!\mathbf{Y},\nonumber \\
R=\Delta\!\mathbf {Y^\mathrm{T} W}\Delta\!\mathbf{Y},\quad
s=\sqrt{\frac{R}{(N-k)}}.\label{def2}
\end{eqnarray}
Correcting $\mathbf{b_e}$ by $\Delta{\mathbf{b}}$ we get the next
solution estimate $\mathbf{b_e}$ and we can repeat the whole
procedure several times. The convergence of accordingly selected LSM
model function is as a rule very swift: after a few steps we state
that $\Delta{\mathbf{b}}\rightarrow \mathbf{0}$,
hence $\mathbf{b=b_e}$.\\[2mm]

{\it 2.3. Uncertainties of parameters and prediction}\\[1mm]

There are at least three reasons why we should estimate the measure
of uncertainty of the found parameters. Firstly, errors of
parameters tell us a lot about the reliability of our results,
secondly uncertainties of parameters would enable to calculate the
uncertainty of the prediction done on the basis of our LSM analysis,
and last but not least above mentioned errors are strictly demanded
by teachers, scientific editors and referees. All LSM instructions
and codes congruently get for uncertainty of the $j$-the parameter
$\delta b_j$ the following relation:
\begin{equation}
\delta b_j=s \sqrt{H_{jj}},\label{chyba}
\end{equation}
where $H_{jj}$ is the $j$-th element in the diagonal of the matrix
\textbf{H}.

It is a question whether $\delta b_j$ really expresses the
uncertainty in the common sense. The response is no, strictly
speaking sometimes yes, but very rarely. It can be demonstrated on
the error of the absolute term in the LSM fit by straight line,
which evidently depends on the choice of the origin of $x$
coordinate.
\begin{figure}
\resizebox{\hsize}{!} {\includegraphics{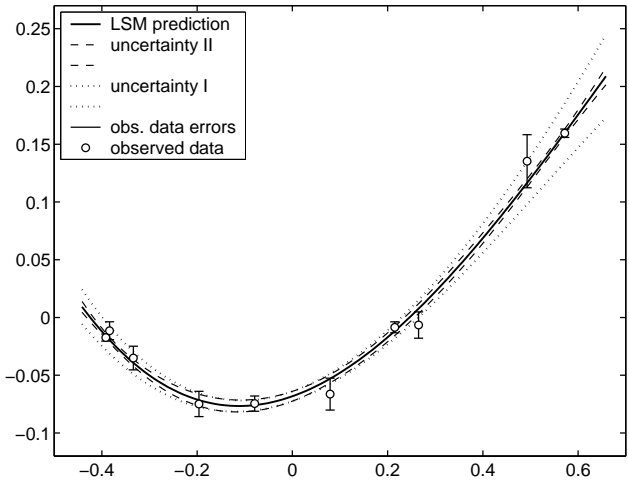}} \caption{The
illustrative figure displays the time dependence of an observed
quantity measured with the accuracy denoted by the abscissa. The
continuous line represents LSM fit by the polynomial of the 3-rd
order (cubic parabola). Expected uncertainties of this prediction
calculated by the formula Eq.\,\ref{blbe} are denoted by dotted
lines, true uncertainties given by Eq.\,\ref{dobre} are signed by
dashed lines.}\label{obr1}
\end{figure}

The suspicion that there is something incorrect in our comprehension
of the true meaning of the quantity $\delta b_j$ defined by
Eq.\,\ref{chyba} will be supported by our attempt use these errors
for the evaluating of the expected uncertainty of the prediction by
the model function for the arbitrarily selected value of $x$:
\begin{equation}
\delta y_p(x)=\sqrt{\sum_{j=1}^k \delta^2
b_j\,f_j^2(x)}=\sqrt{\mathbf{g}(x)\,\mathbf{H_{\mathrm{dg}}\,
g}^\mathrm{T}(x)},\label{blbe}
\end{equation}
where $\mathbf H_{\mathrm{dg}}$ equals to the matrix \textbf{H},
whose all non-diagonal elements has been put zero. $\mathbf{g}(x)$
is the row vector of the gradient of the solution model function
$\mathbf{F}(x,\mathbf{b})$,
$\mathbf{g}(x)=[f_1(x)\,f_2(x)\,\ldots\,f_k(x)]$

The instructive picture Fig.\,\ref{obr1} will show you that this
intuitive relation gives quite inadequate results. Nevertheless, it
can be shown that it is valid formally rather similar relation:
\begin{equation}
\delta y_p(x)=\sqrt{\mathbf{g}(x)\,\mathbf{H\,
g}^\mathrm{T}(x)}.\label{dobre}
\end{equation}

The matrix \textbf{H} is by the definition (see Eq.\,\ref{def1} and
\ref{def2}) a symmetric square $k\!\times\!k$ matrix with $k(k+1)/2$
independent elements. If we want to enable to anybody to compute the
uncertainty of the prediction, we should publish either the whole
matrix \textbf{H} or its non-trivial part at least. Nevertheless,
there is another (more illustrative) possibility: to transform the
model function into the orthogonal one. Then the matrix \textbf{H}
will change in the diagonal one and the uncertainties of parameters
will acquire its standard meaning. It will help you among
other things expertly examine importance of individual terms. \\[2mm]

{\bf 3. Orthogonal LSM models}\\[1mm]

Let us assume that the functional dependence of observed quantities
${y}$ on ${x}$ is well described by the model function which can be
expressed in the form of the linear combination of $k$ basic
functions of $f_j(x)$ with coefficients $b_j$. The found solution
does not change if we use another set of $k$ functions
$\vartheta_j(x)$, which are created as linear combinations of the
basic functions $f_j(x)$. Let us combine them so that the new set of
basic functions $\vartheta_j(x)$ is orthogonal. It means we find the
set of coefficients $\{a_{pj}\}$:
\begin{eqnarray}
\vartheta_p(x)=\sum_{j=1}^k a_{pj}\,f_j(x),\quad \mathrm{so\ that,}\\
\overline{\vartheta_p\,\vartheta_q}=\sum_{i=1}^N
\vartheta_p(x_i)\,\vartheta_q(x_i)\,w_i=0\quad \mathrm{if}\ p\neq
q\label{ortogonalita}
\end{eqnarray}\\
The calculation of linear regression parameters and their
uncertainties is then very simple:
\begin{gather}
b_j=\frac{\sum_{i=1}^N y_i\,\vartheta_j(x_i)\,w_i}{\sum_{i=1}^N
\vartheta_j^2(x_i)\,w_i};\quad\delta b_j=\frac{s}{\sqrt{\sum_{i=1}^N
\vartheta_j^2(x_i)\,w_i}};\nonumber\\
\delta y_p(x)=\sqrt{\sum_{j=1}^k\,\,\delta^2 b_j\,\vartheta_j^2
(x)}.\label{well}
\end{gather}

The set of coefficients $\{a_{pj}\}$ fulfilling constraints
Eq.\,\ref{ortogonalita} is not unique as well as the procedures of
its finding. We recommend to use the following procedure which seems
to us the simplest one:
\begin{gather}
\vartheta_1=f_1;\quad \vartheta_2=f_2-a_{21}\vartheta_1;\nonumber\\
\vartheta_3=f_3-a_{32}\vartheta_2-a_{31}\vartheta_1;\nonumber\\
\vartheta_p(x)=f_p(x)-\sum_{q=1}^{p-1}a_{pq}\,\vartheta_q(x),\label{ortog1}
\end{gather}
where
\begin{equation}
a_{pq}=\frac{\overline{f_p\,\vartheta_q}}{\overline{\vartheta_q^2}}=
\frac{\sum_{i=1}^N f_p(x_i)\,\vartheta_q(x_i)\,w_i}{\sum_{i=1}^N
\vartheta_q^2(x_i)\,w_i}.\label{ortog2}
\end{equation}
The first three orthogonalized terms will be:
\begin{gather}
\vartheta_1(x)=f_1(x); \quad
\vartheta_2(x)=f_2(x)-\frac{\overline{f_2f_1}}{\overline{f_1^2}};\nonumber\\
\vartheta_3(x)=f_3(x)-\frac{\overline{f_3f_2}-\overline{f_3}\
\overline{f_2}}{\overline{f_2^2}-\overline{f_2}^2}\,
f_2(x)\,-\nonumber \\
-\hzav{\frac{\overline{f_3f_1}}{\overline{f_1^2}}-\frac{\overline{f_2f_1}
\,(\overline{f_3f_2}-\overline{f_3}\
\overline{f_2})}{\overline{f_1^2}\zav{\overline{f_2^2}-\overline{f_2}^2}}}
f_1(x).\label{ortog3}
\end{gather}
The explicit expression of successive terms of a set of the
orthogonalized functions is more and more complex, however it is not
very complicated to write an iterative PC code enabling to compute
the formulae for arbitrary number of parameters.\\[2mm]

{\it 3.1. Orthogonal polynomial model}\\[1mm]

The most popular linear regression model (not only in astrophysics)
$F(x,\vec{\beta})$ is:
\begin{equation}
F(x,\vec{\beta})=\sum_{j=1}^k\,\beta_j\,x^{j-1}.\label{PMF}
\end{equation}
The model is known to have a lot uncomfortable properties which
complicate both the calculation and the interpretation of found
results. We should never used it without orthogonalization.

We recommend to put the origin of $x$-coordinates into the center of
gravity of observations: $x\ \rightarrow\ x-\bar x$ before the
application of the orthogonalization procedure. It will result in
the considerable simplification in the form of regression model.
Assuming now $\bar x=0$ the first four orthogonal polynomials are as
follows:
\begin{gather}
\vartheta_1(x)=1;\ \vartheta_2(x)=x;\
\vartheta_3(x)=x^2-\frac{\overline{x^3}}{\overline{x^2}}\,x-\overline{x^2},\nonumber\\
\vartheta_4(x)=x^3-\frac{\overline{x^2}^2\overline{x^3}+\overline{x^3}\,
\overline{x^4}-\overline{x^2}\,\overline{x^5}}{\overline{x^2}^3\!+\overline{x^3}^2\!-
\overline{x^2}\,\overline{x^4}}\,x^2-\nonumber\\
\frac{\overline{x^3}\,\overline{x^5}+\overline{x^2}^2\overline{x^4}-\overline{x^4}^2\!-
\overline{x^3}^2\overline{x^2}}{\overline{x^2}^3\!+\overline{x^3}^2\!-
\overline{x^2}\,\overline{x^4}}\,x\nonumber
-\frac{\overline{x^2}^2\overline{x^5}+\overline{x^3}^3\!-2\overline{x^3}\,
\overline{x^4}}{\overline{x^2}^3\!+\overline{x^3}^2\!-
\overline{x^2}\,\overline{x^4}},
\end{gather}
where,
\begin{equation}
\overline{x^p}=\frac{\sum_{i=1}^N x_i^p\,w_i}{\sum_{i=1}^N w_i}.
\end{equation}
\begin{figure}
\resizebox{\hsize}{!} {\includegraphics{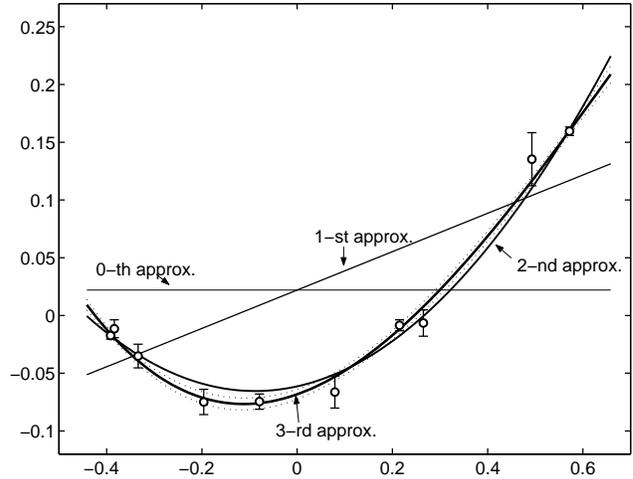}} \caption{The
subsequent approximations of the fit of observed data by orthogonal
polynomial regression.}\label{obr2}
\end{figure}
Fig.\,\ref{obr2} displays the results of subsequent fitting of the
model situation by constant, linear, quadratic and cubic orthogonal
polynomials.

If the data are distributed uniformly in the interval $x_i\in\langle
-\Delta;\,\Delta\rangle$, we can use the transformed Legendre
polynomials (orthogonal on the interval $\langle -1;\,1\rangle$) as
the orthogonal (or quasiorthogonal) LSM model:
\begin{gather}
\vartheta_1=1;\ \vartheta_2=x;\ \vartheta_3=x^2-\frac{\Delta^2}{3};
\ \vartheta_4=x^3-\frac{3\,\Delta^2}{5}\,x;\nonumber\\
\vartheta_5=x^4-\frac{6\,\Delta^2}{7}\,x^2+\frac{3\,\Delta^4}{35};\
\cdots
\end{gather}\\[1mm]

{\it 3.2. Orthogonal sine, cosine model}\\[1mm]

The basic tool for the analysis of cyclic and periodic processes in
astrophysics is the linear regression with the model consisting of
simple periodic functions, the most commonly:
\begin{equation}
F(\varphi,\vec{\beta})=\beta_1+\sum_{j=1}^q\,\beta_{2j}\cos(2\pi
j\varphi)+\beta_{2j+1}\sin(2\pi j\varphi),
\end{equation}
where $\varphi$ is the phase as an independent variable, $q$ is the
order of set of harmonic functions. The model need not contain all
of functions, it might be limited e.g. only to even functions etc.

In the case that the observations are spread over the whole cycle
more or less uniformly, it is not needed to do any
orthogonalization, because all functions are orthogonal itself. In
the opposite case we should do orthogonalization e.g. by the
procedure
described by Eq.\,\ref{ortog1} and Eq.\,\ref{ortog2}.\\[2mm]

{\bf 4. Conclusions}\\[1mm]

We displayed the benefits of consequential usage of orthogonal LSM
model functions with the emphasis on the polynomial regression as
the chief tool of astrophysical data processing. Orthogonal models
enable to give the true sense to errors of found parameters and
easily compute estimates for uncertainties of the prediction. The
orthogonality of the models removes the bad conditioning of the
solved systems of equations and help us to obtain results not
affected by computational errors. We recommend to use them always,
compulsorily in the case of polynomial regression.

It is demanding to use new methods of variable stars data processing
which enable us better exploit information hidden in observations.
Endeavor connected with mastering of them will return in new subtle
discoveries and revealing.

Matrix calculus, true using of weights, advanced principal component
analysis, factor analysis, robust regression, creation and usage of
orthogonal models and several other processing techniques should
appertain to compulsory outfit of each variable stars' observer of
the 21st century.\\[2mm]

{\it Acknowledgements.} This work was supported by grants GA\,\v{C}R
205/06/0217, and MVTS \v{C}RSR 10/15. The author is indebted to
prof.~Izold Pustylnik and Dr.~Miloslav Zejda for careful and
critical reading of the manuscript and suggestions which
considerable improved the article.
\\[3mm]
\indent
{\bf References\\[2mm]}
Mikul\'{a}\v{s}ek Z.: 2007, {\it Astron.\,Astrophys.\,Trans.,} {\bf
26}, 63.
\end{document}